\documentclass[nohyper,12pt,letterpaper]{JHEP}
\usepackage{epsfig}
\newfont{\frak}{eufm10 scaled 1200}

\newfont{\Bbb}{msbm10 scaled 1200}
\newcommand{\mathbb}[1]{\mbox{\Bbb #1}}
\DeclareSymbolFont{AMSa}{U}{msa}{m}{n}
\DeclareSymbolFont{AMSb}{U}{msb}{m}{n}
\let\Box\relax
\DeclareMathSymbol{\Box}{\mathord}{AMSa}{"03}

\title{Noncummutative Decrumpling Inflation and Running of the Spectral
Index}

\author{Forough Nasseri\\
Department of Physics, Tarbiat Moallem University of Sabzevar,\\
P.O.Box 397, Sabzevar, Iran\\
and\\
Khayyam Planetarium, P.O.Box 844, Neishabour, Iran\\
E-mail: \email{nasseri@fastmail.fm}}

\author{S. A. Alavi\\
Department of Physics, Tarbiat Moallem University of Sabzevar,\\
P.O.Box 397, Sabzevar, Iran\\
E-mail: \email{alavi@sttu.ac.ir}}

\abstract{We present a new inflation model, known as noncommutative
decrumpling inflation, in which space has noncommutative
geometry with time variability of the number of spatial dimensions.
Within the framework of noncommutative decrumpling inflation,
we compute both the spectral index and its running.
Our results show the effects of both time variability of
the number of spatial dimensions and noncommutative geometry
on the spectral index and its running. Two classes of examples
have been studied and comparisons made with the standard slow-roll
formulae. We conclude that the effects of noncommutative
geometry on the spectral index and its running are much smaller than
the effects of time variability of spatial dimensions.}

\keywords{Inflation; noncommutative geometry; extra dimensions.\\
Report number: STTU-Phys-02-2004}

\date{today}
\preprint{STTU-Phys-02-2004}
\begin{document}

\section{Introduction}

In nature there are three spatial dimensions and one time dimension.
The idea of having space-time dimensions other than $3+1$ goes
back to Kaluza-Klein theory. This concept has been generalized to
string theory with space dimension more than three but still an
integer and a constant.

We here present a new inflation model, known as noncommutative
decrumpling inflation model in which not only the number of spatial
dimension has a dynamical behavior and decreases but also space
has a noncommutative geometry.

Our motivation to study noncommutative decrumpling inflation is
to investigate cosmological implications of both time
variability of the number of spatial dimensions and noncommutativity of
space. To do so, we compute the spectral index
and its running within the framework of noncommutative decrumpling
inflation. For more details about a model Universe with time variable
space dimensions (TVSD), known as TVSD or decrumpling model,
see Refs. [1-7].

This article is a follow-up paper of Ref. [4] about TVSD or decrumpling
chaotic inflation model. To do this research an important conceptual
issue is properly dealt with the meaning of time variability of the
number of spatial dimensions. In Ref. [2], this conceptual issue has been
discussed in detail.

Although time variability of spatial dimensions have not been firmly
achieved in experiments and theories, such dynamical behavior of the
spatial dimensions should not be ruled out in the context of cosmology
and astroparticle physics.

Here, we will be concerened with the approaches proposed in the pioneer
paper\cite{1} where the cosmic expansion of the Universe is named
decrumpling expansion and is due to decrease of the number of spatial
dimensions.
The most important difference between decrumpling model and other
attempts about the time evolution of spatial dimension is that in this
model the number of extra spatial dimensions changes with time while
in other theories the size of extra spatial dimensions is a dynamical
parameter \cite{{8},{9}}. Based on time variability of the size of
spatial dimensions it has been reported\cite{9} that
the present rate of change of the mean radius of any additional spatial
dimensions to be less than about $10^{-19} {\rm yr^{-1}}$. It is worth
mentioning that the result of Ref. [9] is based on dynamical behavior
of the size of extra spatial dimensions while in decrumpling model we
take the size of extra spatial dimensions to be constant and the number
of spatial dimensions decreases continuously as the Universe expands.
The present rate of time variation of the number of the spatial dimensions
in decrumpling or TVSD model is about $10^{-13} {\rm yr^{-1}}$.

Another subject which lately has attracted much attention
is noncommutative spaces. It is generally believed that the picture of
space-time as a manifold should break down at very short distances of
the order of the Planck length. Field theories on noncommutative
spaces may play an important role in unraveling the properties of
nature at the Planck scale. It has been shown that the noncommutative
geometry naturally appears in string theory with a non zero antisymmetric
B-field \cite{10}.

Besides the string theory arguments the noncommutative field theories
by themselves are very interesting. In a noncommutative space-time
the coordinates operators satisfy the commutative relation
\begin{equation}
\label{1}
[\hat{x}^\mu,\hat{x}^\nu]=i\theta^{\mu\nu},
\end{equation}
where $\hat{x}$ are the coordinate operators and $\theta^{\mu\nu}$
is an antisymmetric tensor of dimension of (length)$^2$.
Generally noncommutative version of a field theory is obtained by
replacing the product of the fields appearing in the action by the star
product
\begin{equation}
\label{2}
(f \star g)(x)=\exp \left( \frac{i}{2} \theta^{\mu\nu} \frac{\partial}
{\partial x^{\mu}}\frac{\partial}{\partial y^{\nu}}\right)
f(x)g(y)|_{x=y}
\end{equation}
where $f$ and $g$ are two arbitrary functions which we assume to be
infinitely differentiable.

In recent years there have been a lot of works devoted to the study
of noncommutative field theory or noncommutative quantum mechanics
and possible experimental consequences of extensions of the standard
formalism \cite{11}-\cite{19}.
In the last few years there has been also a growing interest in
probing the space-space noncommutativity effects on cosmological
observations \cite{20}-\cite{25}.

We will use the natural units system that sets $k_B$, $c$, and $\hbar$
all equal to one, so that $\ell_P=M_P^{-1}=\sqrt{G}$.
To read easily this article we also use the notation $D_t$ instead of
$D(t)$ that means the space dimension $D$ is as a function of time.

The plan of this article is as follows. In section 2, we give a brief
review of decrumpling model in commutative spaces.
In section 3, we first present the explicit and
general formulae for the spectral index and its running
within the framework of noncommutative decrumpling inflation
and then apply them to two classes of examples of the inflaton potential.
Finally, we discuss our results and conclude in section 4.

\section{Review of Decrumpling or TVSD Model}

Decrumpling model is based on the assumption that the
basic blocks of the space-time are fractaly
structured \cite{{1},{2}}.
In the pioneer paper \cite{1}
the spatial dimension of the Universe was considered as a continuous
time dependent variable. As the Universe expands, its spatial dimension
decreases continuously, thereby generating what has been named a
decrumpling Universe. Then this model has been overlooked and the quantum
cosmological aspects, as well as, a possible test theory for studying
time evolution of Newton's constant have also been
discussed \cite{{2},{3}}. Chaotic inflation in decrumpling model 
and its dynamical solutions have also been studied \cite{{4},{5}}.

The concept of decrumpling expansion of the Universe
is inspired by the idea of decrumpling coming from polymer
physics \cite{{1},{2}}. In this model the fundamental building blocks of
the Universe are like cells with arbitrary dimensions having in
each dimension a characteristic size $\delta$ which maybe of the order of
the Planck length ${\mathcal O}$($10^{-33}$ cm) or even smaller
so that the minimum physical radius of the Universe is $\delta$.
These ``space cells'' are embedded in a ${\mathcal D}$ space, where
${\mathcal D}$ may be up to infinity. 
Therefore, the space dimensions of the
Universe depend on how these fundamental cells are configured in this
embedding space. The Universe may have begun from a very crumpled
state having a very high dimension ${\mathcal{D}}$ and a size $\delta$,
then have lost dimension through a uniform decrumpling which we see
like a uniform expansion. The expansion of space, being now understood like
a decrumpling of cosmic space, reduces the space-time dimension continuously
from ${\mathcal{D}}+1$ to the present value $D_0+1$.
In this picture, the Universe can have any space dimension.
As it expands, the number of spatial dimensions decreases continuously.
The physical process that causes or necessitates such a decrease in
the number of spatial dimensions comes from how these fundamental cells
are embedded in a ${\mathcal{D}}$ space.

As an example, take a limited number of small three-dimensional beads.
Depending on how these beads are embedded in space they can configure to a
one-dimensional string, two-dimensional sheet, or three-dimensional sphere.
This is the picture we are familiar with from the concept of crumpling
in polymer physics where a crumpled polymer has a dimension more
than one. Or take the picture of a clay which can be like a
three-dimensional sphere, or a two-dimensional sheet, or even a
one-dimensional string, a picture based on the theory of fluid membranes.

While it is common to make {\it ad hoc} assumptions in cosmological model
building in the absence of a complete theory of quantum gravity,
some of the particular ingredients of the model owe their physical basis
perhaps more to polymer physics than to cosmology. 
Progress with decrumpling model can only be made if there is a breakthrough
in terms of finding a natural mechanism for varying the number of
spatial dimensions in some alternative fashion to that which is considered here.

For more details about motivation for choosing this model and its
technical and conceptual formalism, see Ref. [2].

\subsection{Relation between the effective space dimension
$D(t)$ and characteristic size of the Universe $a(t)$}
\noindent
Assume the Universe consists of a fixed number $\mathbb{N}$ of universal
cells having a characteristic length $\delta$ in each of their
dimensions. The volume of the Universe at the time $t$ depends
on the configuration of the cells. It is easily seen that \cite{{1},{2}}
\begin{equation}
\label{3}
{\rm vol}_{D_t}({\rm cell})={\rm vol}_{D_0}({\rm cell})\delta^{D_t-D_0},
\end{equation}
where the $t$ subscript in $D_t$ means that $D$ to be as a function
of time, i.e. $D(t)$. In previous references \cite{1}-\cite{7} the
notation of $D_t$ did not use for $D(t)$. 

Interpreting the radius of the Universe, $a$, as the radius of
gyration of a crumpled ``universal surface'',
the volume of space can be written \cite{{1},{2}}
\begin{eqnarray}
\label{4}
a^{D_t}&=&\mathbb{N}\,{\rm vol}_{D_t}({\rm cell})\nonumber\\
   &=&\mathbb{N}\,{\rm vol}_{D_0}({\rm cell}) \delta^{{D_t}-D_0}\nonumber\\
   &=&{a_0}^{D_0} \delta^{{D_t}-D_0}
\end{eqnarray}
or
\begin{equation}
\label{5}
\left( \frac{a}{\delta} \right)^{D_t}=
\left( \frac{a_0}{\delta} \right)^{D_0} = e^C,
\end{equation}
where $C$ is a universal positive constant. Its value has a strong
influence on the dynamics of space-time, for example on the dimension
of space, say, at the Planck time. Hence, it has physical and cosmological
consequences and may be determined by observation. The zero subscript in
any quantity, e.g. in $a_0$ and $D_0$, denotes its present value.
We coin the above relation as a``dimensional constraint" which relates
the ``scale factor" of decrumpling model to the spatial dimension.
We consider the comoving length of the Hubble
radius at present time to be equal to one. So the interpretation of the
scale factor as a physical length is valid.
The dimensional constraint can be written in this form \cite{{1},{2}}
\begin{equation}
\label{6}
\frac{1}{D_t}=\frac{1}{C}\ln \left( \frac{a}{a_0} \right) + \frac{1}{D_0}.
\end{equation}
It is seen that by the expansion of the Universe, the space
dimension decreases.
Time derivative of (\ref{5}) or (\ref{6}) leads to
\begin{equation}
\label{7}
{\dot D}_t=-\frac{D_t^2 \dot{a}}{Ca}.
\end{equation}
It can be easily shown that the case of constant space dimension
corresponds to when $C$ tends to infinity. In other words,
$C$ depends on the number of fundamental cells. For $C \to +\infty$,
the number of cells tends to infinity and $\delta\to 0$.
In this limit, the dependence between the space dimensions and
the radius of the Universe is removed, and consequently we
have a constant space dimension \cite{{1},{2}}.

\subsection{Physical meaning of $D_P$}

We define $D_{P}$ as the space dimension of the Universe when
the scale factor is equal to the Planck length $\ell_{P}$.
Taking $D_0=3$ and the scale of the Universe today to be the present
value of the Hubble radius $H_0^{-1}$ and the space dimension at the
Planck length to be $4, 10,$ or $25$, from Kaluza-Klein and superstring
theory, we can obtain from (\ref{5}) and (\ref{6}) the corresponding
value of $C$ and $\delta$
\begin{eqnarray}
\label{8}
\frac{1}{D_P}&=&\frac{1}{C} \ln \left( \frac{\ell_P}{a_0}
\right) + \frac{1}{D_0}
= \frac{1}{C} \ln \left( \frac{\ell_P}{{H_0}^{-1}} \right) +\frac{1}{3},\\
\label{9}
\delta &=& a_0 e^{-C/D_0}= H_0^{-1} e^{-C/3}.
\end{eqnarray}
In Table 1, values of $C$, $\delta$ and also ${\dot D}_t|_0$ for some
interesting values of $D_P$ are given.\cite{2}-\cite{5}.
These values are calculated by assuming $D_0=3$ and
$H_0^{-1}=3000 h_0^{-1} {\rm Mpc}=9.2503 \times 10^{27} h_0^{-1}{\rm cm}$,
where we take $h_0=1$, see Ref. [2].
\TABLE{Table 1. Values of $C$ and $\delta$ for some values of $D_P$.
Time variation of space dimension today has also been calculated
in terms of ${\rm yr}^{-1}$, see Ref. [2].
\begin{tabular}{cccc}
$D_P$ & $C$ & $\delta\,({\rm cm})$ & ${\dot D}_t|_0\,({\rm yr}^{-1})$\\ \hline\hline
$3$       & $+\infty$ & $0$ & $0$ \\ \hline
$4$       & $1678.8$  & $8.6 \times 10^{-216}$ & $-5.5 \times 10^{-13} $ \\ \hline
$10$      & $599.6$   & $1.5 \times 10^{-59}$  & $-1.5 \times 10^{-12} $ \\ \hline
$25$      & $476.9$   & $8.3 \times 10^{-42}$  & $-1.9 \times 10^{-12} $ \\ \hline
$+\infty$ & $419.7$   & $\ell_P$               & $-2.2 \times 10^{-12} $ \\ \hline\hline
\end{tabular}}

\section{Running of the Spectral Index in Noncommutative Decrumpling
Inflation}

In this section, we present explicit and general formulae for the
spectral index and its running within the framework of noncommutative
decrumpling inflation.
To do so, we first present the formalism of decrumpling chaotic
inflation \cite{4} in commutative spaces and then apply them
in the context of noncommutative spaces.
For the purposes of illustration, we apply our results for two
classes of examples of the inflaton potential.

\subsection{Decrumpling chaotic inflation}

Inflation has been studied in the framework of decrumpling or TVSD
model. The crucial equations are \cite{4}
\begin{eqnarray}
\label{10}
&&H^2 = \left( \frac{\dot a}{a} \right)^2 =
\frac{16 \pi}{D_t(D_t-1)M_P^2} \left( \frac{1}{2} {\dot\phi}^2 +
V(\phi) \right) -\frac{k}{a^2},{\mbox{Friedmann equation}},\\
\label{11}
&&\ddot{\phi}+D_tH\dot\phi+{\dot D}_t \dot\phi \left( \ln \frac{a}{a_0} +
\frac{d \ln V_{D_t}}{dD_t} \right) = -V'(\phi),\;\;\;\;\;
{\mbox{Fluid equation}},
\end{eqnarray}
where $V_{D_t}$ is the volume of the space-like sections \cite{2}
\begin{eqnarray}
\label{12}
V_{D_t} &=& \cases {\frac{2 \pi^{(D_t+1)/2}}{\Gamma[(D_t+1)/2]},
               & if $\;k=+1,\;\;\;{\mbox{closed decrumpling model,}}$ \cr
               \frac{\pi^{(D_t/2)}}{\Gamma(D_t/2+1)}{\chi_c}^{D_t},
               & if $\;k=0,\;\;\;{\mbox{flat decrumpling model,}}$ \cr
               \frac{2\pi^{(D_t/2)}}{\Gamma(D_t/2)}f(\chi_c),
               & if $\;k=-1,\;\;\;{\mbox{open decrumpling model.}}$ \cr}
\end{eqnarray}
These volumes of space-like sections are valid even in the case
of constant $D$-space \cite{2}.
Here $\chi_C$ is a cut-off and $f(\chi_c)$ is a function
thereof \cite{2}.

Using the slow-roll approximation in decrumpling model \cite{4}
\begin{eqnarray}
\label{13}
{\dot{\phi}}^2 &\ll& V(\phi),\\
\label{14}
{\ddot{\phi}} &\ll& D_tH{\dot \phi},\\
\label{15}
-{\dot H} &\ll& H^2
\end{eqnarray}
and 
\begin{equation}
\label{16}
{\dot D}_t \left( \ln \frac{a}{a_0} +
\frac{d \ln V_{D_t}}{d D_t} \right) \ll D_t H,
\end{equation}
Eqs. (\ref{10}) and (\ref{11}) can be rewritten for a flat decrumpling
model, i.e. $k=0$, in the simpler set
\begin{eqnarray}
\label{17}
H^2 &\simeq& \frac{16 \pi V(\phi)}{D_t(D_t-1) M_P^2},\\
\label{18}
D_t H \dot{\phi} &\simeq& - V'(\phi).
\end{eqnarray}
Note that the slow-roll condition (\ref{16}) has not been considered
in Ref. [4]. The validity of this
condition is obvious by regarding Eq. (\ref{7}). Substituting (\ref{7})
in (\ref{17}), dynamics of the spatial dimension is given by \cite{4}
\begin{equation}
\label{19}
{{\dot D}_t}^2 \simeq  \frac{16 \pi D_t^3 V(\phi)}{C^2 (D_t -1) M_P^2}.
\end{equation}
During inflation, $H$ is slowly varying in the sence that its change per
Hubble time $\epsilon \equiv -{\dot H}/{H^2}$ is less than one.
The slow-roll condition $|\eta| \ll 1$ is actually a consequences of the
condition $\epsilon \ll 1$ plus the slow-roll approximation
$D_t H {\dot \phi} \simeq - V'(\phi)$. Deferentiating (\ref{18}) one finds
\begin{equation}
\label{20}
\frac{\ddot \phi}{H \dot{\phi}} = \epsilon - \eta + \frac{D_t}{C},
\end{equation}
where the slow-roll parameters in decrumpling model are defined by
\begin{eqnarray}
\label{21}
\epsilon &\equiv& \frac{(D_t-1) M_P^2}{32 \pi}
\left( \frac{V'}{V} \right)^2,\\
\label{22}
\eta &\equiv& \frac{(D_t-1) M_P^2}{16 \pi} \left( \frac{V''}{V} \right).
\end{eqnarray}
It should be emphasized that the slow-roll parameters in decrumpling model
as presented in Ref. [4] are different from those given in (\ref{21}) and
(\ref{22}).
This difference is due to the slow-roll condition (\ref{16}) which has
not been considered in Ref. [4]. Furthermore, in the constant
$D$-space, the slow-roll parameters (\ref{21}) and (\ref{22}) are also
valid by substituting $D_t$ by $D$, see Ref. [4]. 

\subsection{Explicit formulae for running in noncommutative decrumpling
inflation}

The density perturbation in noncommutative decrumpling inflation can be
expressed by \cite{24}
\begin{equation}
\label{23}
\delta_H^2=\left(\frac{H}{2\pi}\right)^2 \left( 1-\frac{3}{32} \frac{H^4}
{\Lambda^4} \sin^2\vartheta \right)^2 \left( \frac{H}{\dot{\phi}} \right)^2,
\end{equation}
where $\vartheta$ is the angle between the comoving wave-number $k$ and the
third axis and $\Lambda^{-1}$ is the noncommutativity length
scale \cite{{24},{25}}.

The density perturbation in noncommutative decrumpling
inflation can be determined by substituting (\ref{17}) and (\ref{18})
in (\ref{23}) 

\begin{equation}
\label{24}
\delta_H^2=\frac{9216 \pi V^3}{D_t^3 (D_t-1)^3 M_P^6 V'^2}
\left( 1 - \frac{3}{32} \frac{H^4}{\Lambda^4} \sin^2
\vartheta \right)^2.
\end{equation}
This expression is evaluated at the horizon crossing time when
$k=aH$. Since the value of Hubble constant does not change too much during
inflationary epoch, we can obtain $dk = Hda$ and
$d\ln k = Hdt = da/a$ . Using the slow-roll
condition in decrumpling inflation
\begin{equation}
\label{25}
\frac{d}{d \ln k} = - \frac{V'}{D_t H^2} \frac{d}{d \phi},
\end{equation}
and also the dimensional constraint of the model we have
\begin{eqnarray}
\label{26}
\frac{dD_t}{da}=-\frac{D_t^2}{Ca},\\
\label{27}
\frac{dD_t}{d \ln k} = - \frac{D_t^2}{C},\\
\label{28}
\frac{dD_t}{d\phi}= \frac{D_t^3 H^2}{V' C}.
\end{eqnarray}
After a lengthy but straightforward calculation
by using Eqs. (\ref{17}), (\ref{18}) and (\ref{24})-(\ref{28}) we find
\begin{eqnarray}
\label{29}
n_S-1 &\equiv& \frac{d \ln \delta_H^2}{d \ln k}\nonumber\\
&=&-6 \epsilon +2 \eta + \frac{6 \pi V'^2 \sin^2 \vartheta}
{D_t^2 (D_t-1) \Lambda^4 M_P^2} \left( 1- \frac{3}{32} \frac{H^4}{\Lambda^4}
\sin^2 \vartheta \right)^{-1} \nonumber\\
&\times& \bigg[ 1 - \frac{16 \pi D_t (2D_t-1)}
{C(D_t-1)^2 M_P^2} \left( \frac{V}{V'} \right)^2 \bigg] +
\frac{3D_t(2D_t-1)}{C(D_t-1)},
\end{eqnarray}
where $\epsilon$ and $\eta$ are given in (\ref{21}) and (\ref{22}).
To calculate the running we use the following expressions
\begin{eqnarray}
\label{30}
\frac{d \epsilon}{d \ln k} &=& - \frac{D_t^2}{C(D_t-1)}\epsilon -
2 \epsilon \eta + 4 \epsilon^2,\\
\label{31}
\frac{d \eta}{d \ln k} &=& -\frac{D_t^2}{C(D_t-1)} \eta +
2 \epsilon \eta- \xi,
\end{eqnarray}
where the third slow-roll parameter is defined by
\begin{equation}
\label{32}
\xi \equiv \frac{(D_t-1)^2M_P^4}{(16 \pi)^2}\left(\frac{V'V'''}{V^2}\right).
\end{equation}
Using above equations, running in noncommutative decrumpling inflation
has this explicit expression
\begin{eqnarray}
\label{33}
&&\frac{d n_S}{d \ln k}= 16 \epsilon \eta
- 24 \epsilon^2 - 2 \xi
+\frac{6 D_t^2}{C(D_t-1)} \epsilon-\frac{2D_t^2}{C(D_t-1)} \eta
-\frac{3D_t^2(2D_t^2-4D_t+1)}{C^2(D_t-1)^2}\nonumber\\
&&-\frac{6 \sin^2 \vartheta}{D_t^2(D_t-1)\Lambda^4}
\left( 1- \frac{3}{32}
\frac{H^4}{\Lambda^4} \sin^2 \vartheta \right)^{-2}
\bigg[ \frac{3 \pi^2 V'^2 \sin^2 \vartheta}{M_P^4 \Lambda^4}
\bigg( \frac{V'^2}{D_t^2(D_t-1)}\nonumber\\
&&-\frac{32 \pi (2D_t-1)V^2}{CD_t(D_t-1)^3M_P^2}
+\frac{256 \pi^2 (2D_t-1)^2V^4}{C^2 (D_t-1)^5 M_P^4 V'^2} \bigg)
+ \left( 1-\frac{3}{32} \frac{H^4}{\Lambda^4} \sin^2 \vartheta \right)
\nonumber\\
&&\times \bigg( \frac{V'^2 V'' (D_t-1)}{8 V}
-\frac{\pi V'^2 D_t(7D_t-4)}{C(D_t-1)M_P^2}
-\frac{16 \pi^2 V^2 D_t^2 (6D_t^2-2D_t-1)}
{C^2 (D_t-1)^3 M_P^4} \bigg) \bigg].
\end{eqnarray}
Let us now use (\ref{29}) and (\ref{33})  for two potentials of the inflation.
We first study $V(\phi)=m^2 \phi^2/2$ and then $V(\phi)=\lambda \phi^4$.

\subsection{The first example: $V(\phi)=\frac{1}{2}m^2\phi^2$}

For the purposes of illustration, we now consider the potential
$V(\phi)=m^2 \phi^2/2$

\begin{eqnarray}
\label{34}
\epsilon&=&\eta=\frac{(D_t-1)M_P^2}{8 \pi \phi^2},\\
\label{35}
\xi&=&0.
\end{eqnarray}
Using the definition of e-folding in decrumpling inflation {\cite4}
\begin{equation}
\label{36}
{\mathcal{N}}=-\frac{16\pi}{M_P^2} \int_{\phi}^{\phi_f}
\frac{V}{(D_t-1)V'}d\phi,
\end{equation}
we have
\begin{equation}
\label{37}
{\mathcal{N}}=\frac{4\pi}{(D_t-1)M_P^2}(\phi^2-\phi_f^2),
\end{equation}
where $\phi_f$ is the value of the inflaton field at the end of
inflation. Note that in the integral of (\ref{36}),
$D_t$ is as a function of the inflaton field. To integrate in (\ref{36})
we take $D_t$ to be independent on the inflaton field because
the relationship between $D_t$ and $\phi$ in decrumpling inflation is
too weak \cite{5}. For this reason, our approximation is appropriate for
integration of Eq. (\ref{36}).

To obtain e-folding as a function of the inflaton field, we must obtain
$\phi_f$. From $\epsilon=\eta=1$ we get
\begin{equation}
\label{38}
\phi_f=\sqrt{\frac{D_t-1}{8\pi}}M_P.
\end{equation}
From (\ref{37}) and (\ref{38}), we have
\begin{equation}
\label{39}
\phi^2=\frac{(D_t-1)M_P^2}{8\pi}(2{\mathcal{N}}+1).
\end{equation}
With no loss of generality, we take $\vartheta=\frac{\pi}{2}$.
So substituting $\sin \vartheta=1$ in the above equations leads us to
\begin{eqnarray}
\label{40}
n_S-1&=&-\frac{4}{(2{\mathcal{N}}+1)}+
\frac{3(2{\mathcal{N}}+1)m^4}{4D_t^2\Lambda^4} \left( 1-
\frac{3(2{\mathcal{N}}+1)^2m^4}{32D_t^2\Lambda^4}\right)^{-1}\nonumber\\
&\times&\bigg[ 1 -\frac{D_t(2D_t-1)(2{\mathcal{N}}+1)}{2C(D_t-1)}\bigg]+
\frac{3D_t(2D_t-1)}{C(D_t-1)},\\
\label{41}
\frac{d n_S}{d \ln k}&=& -\frac{8}{(2{\mathcal{N}}+1)^2}-
\frac{4D_t^2}{C(D_t-1)(2{\mathcal{N}}+1)}-
\frac{3D_t^2(2D_t^2-4D_t+1)}{C^2(D_t-1)^2}\nonumber\\
&-&\frac{3m^4}{2D_t^2(D_t-1)\Lambda^4}\left( 1 -
\frac{3(2{\mathcal{N}}+1)^2m^4}{32D_t^2
\Lambda^4} \right)^{-2} \bigg[ \frac{3 m^4 (D_t-1)^2(2{\mathcal{N}}+1)^2}
{16 \Lambda^4}\nonumber\\
&\times& \left( \frac{1}{D_t^2(D_t-1)} -
\frac{(2D_t-1)(2{\mathcal{N}}+1)}{CD_t(D_t-1)^2}
+\frac{(2D_t-1)^2(2{\mathcal{N}}+1)^2}{4C^2(D_t-1)^3} \right)\nonumber\\
&+& \left( 1- \frac{3(2{\mathcal{N}}+1)^2m^4}{32 D_t^2 \Lambda^4}\right)
\bigg( D_t-1-\frac{D_t(7D_t-4)(2{\mathcal{N}}+1)}{2C}\nonumber\\
&-&\frac{D_t^2(6D_t^2-2D_t-1)(2{\mathcal{N}}+1)^2}
{4C^2(D_t-1)} \bigg) \bigg].
\end{eqnarray}
The point is that the general formulae (\ref{40}) and (\ref{41})
in terms of unknown value of $\vartheta$ could be obtained by
substituting $\Lambda^2$ with $\Lambda^2/\sin\vartheta$.

\subsection{The second example: $V(\phi)=\lambda \phi^4$}

For the second example, we study $V(\phi)=\lambda \phi^4$.
The slow-roll parameters are
\begin{eqnarray}
\label{42}
\epsilon &=& \frac{(D_t-1)M_P^2}{2\pi \phi^2},\\
\label{43}
\eta&=&\frac{3(D_t-1)M_P^2}{4\pi\phi^2},\\
\label{44}
\xi&=&\frac{3(D_t-1)^2M_P^4}{8\pi^2\phi^4}.
\end{eqnarray}
Using the definition of e-folding in inflation, we have
\begin{equation}
\label{45}
{\mathcal{N}}=\frac{2\pi}{M_P^2(D_t-1)}\left(\phi^2-\phi_f^2 \right).
\end{equation}
To obtain the e-folding number as a function of the inflaton field,
we must obtain the value of the inflaton field at the end of inflation.
From $\epsilon=1$ we get
\begin{equation}
\label{46}
\phi_f=\sqrt{\frac{D_t-1}{2\pi}}M_P.
\end{equation}
Assuming $\eta=1$ and $\xi=1$,
we obtain
\begin{equation}
\label{47}
\phi_f=\sqrt{\frac{3(D_t-1)}{4\pi}}M_P
\end{equation}
and
\begin{equation}
\label{48}
\phi_f=\left( \frac{3(D_t-1)^2}{8\pi^2} \right)^{1/4}M_P,
\end{equation}
respectively. These values of $\phi_f$ based on the condition
$\eta=1$ and $\xi=1$ are larger than $\phi_f$ arisen from $\epsilon=1$.
We here take the condition $\epsilon=1$ by itself is a true condition to
obtain $\phi_f$. From (\ref{45}) and (\ref{46}) we have
\begin{equation}
\label{49}
\phi^2=\frac{(D_t-1)({\mathcal{N}}+1)M_P^2}{2\pi}.
\end{equation}
Again we take $\vartheta=\pi/2$. Substituting  $\,\sin \vartheta=1$ in
(\ref{29}) and (\ref{33}) and using (\ref{42}), (\ref{43}) and
(\ref{44}) we obtain the main formulae for the spectral index and its
running
\begin{eqnarray}
\label{50}
&&n_S-1 = - \frac{3}{({\mathcal{N}}+1)} + \frac{12
\lambda^2 M_P^4 (D_t-1)^2 ({\mathcal{N}}+1)^3}
{\pi^2 D_t^2 \Lambda^4}\nonumber\\
&& \times \left(1-\frac{3 \lambda^2 M_P^4 (D_t-1)^2({\mathcal{N}}+1)^4}
{2 \pi^4 D_t^2 \Lambda^4} \right)^{-1}
\bigg[ 1- \frac{D_t(2D_t-1)({\mathcal{N}}+1)}{2C(D_t-1)} \bigg]\nonumber\\
&&+\frac{3D_t(2D_t-1)}{C(D_t-1)},\\
\label{51}
&&\frac{dn_S}{d\ln k}=-\frac{3}{({\mathcal{N}}+1)^2}+
\frac{3D_t^2}{C(D_t-1)({\mathcal{N}}+1)}
-\frac{24 \lambda^2M_P^4(D_t-1)({\mathcal{N}}+1)^2}
{\pi^2D_t^2 \Lambda^4}\nonumber\\
&&\times \left( 1 - \frac{3 \lambda^2 M_P^4 (D_t-1)^2
({\mathcal{N}}+1)^4}{2\pi^2D_t^2
\Lambda^4}\right)^{-2} \bigg\{ \frac{3\lambda^2 M_P^4(D_t-1)^4
({\mathcal{N}}+1)^4}
{\pi^2 \Lambda^4}\nonumber\\
&&\times\bigg[ \frac{1}{D_t^2(D_t-1)} - \frac{(2D_t-1)({\mathcal{N}}+1)}
{CD_t(D_t-1)^2}+\frac{(2D_t-1)^2({\mathcal{N}}+1)^2}{4C^2(D_t-1)^3}
\bigg]\nonumber\\
&&+\left( 1 - \frac{3 \lambda^2 M_P^4 (D_t-1)^2
({\mathcal{N}}+1)^4}{2 \pi^2 \Lambda^4
D_t^2} \right) \bigg[ \frac{3(D_t-1)}{2} - \frac{D_t(7D_t-4)
({\mathcal{N}}+1)}{2C}\nonumber\\
&&-\frac{D_t^2(6D_t^2-2D_t-1)({\mathcal{N}}+1)^2}{4C^2(D_t-1)}
\bigg] \bigg\} -\frac{3D_t^2 (2D_t^2-4D_t+1)}{C^2(D_t-1)^2}.
\end{eqnarray}

\section{Conclusions}

We have studied the effects of both space-space noncommutative
geometry and time variability of spatial dimensions on
the spectral index and its running, and obtained the correction terms
to the standard formulae. If there exists any time variability of
spatial dimensions and noncommutativity of space in nature, as it seems
to emerge from different theories and arguments, its implications should
appear in the spectral index and its running. We presented the general
features of our formalism and applied it to two specific potentials of
chaotic inflation. The results show that the terms arising from 
noncommutativity of space depend on the space-space noncommutativity
length scale and those arising from time variability of spatial
dimensions depend on the universal parameter of decrumpling model.

For the purposes of illustration, we present the formulae for the spectral
index and its running in noncommutative decrumpling inflation for two
classes of examples of the inflaton potentials. The first one
is $m^2 \phi^2/2$ and the second $\lambda \phi^4$.

Since the upper value of the universal parameter, $C$, of decrumpling
model is about $1680$, the correction terms arising from time variability
of spatial dimensions, being of the order of ${\mathcal{O}}(1/C)$
or its powers of $2$ or so, are of the order of $10^{-3}$ or less.
On the other hand, considering the noncommutativity length scale
$\Lambda^{-1}$, being of the order of the Planck length, so we obtain
$M_P^4/\Lambda^4$ of the order of unity, ${\mathcal{O}}(1)$. This point
simplifies the formulae of the spectral index and its running, see Eqs.
(\ref{50}) and (\ref{51}).
We know the mass of the inflaton field to
be $m \simeq 1.21 \times 10^{-6} M_P$, the coupling constant
$\lambda \simeq 10^{-15}$ and the e-folding number
${\mathcal{N}}\simeq 60$. Regarding these values of $m$, $\lambda$ and
${\mathcal{N}}$, and comparing the correction terms due to time variability
of spatial dimensions with those due to space-space
noncommutative geometry, we conclude that the effects of noncommutative
spaces on the spectral index and its running are much smaller than
the effects of time variability of spatial dimensions.
When $C \to +\infty$ and $D_t=D_0=3$ our results in this article
reproduce the standard formulae in noncommutative three-space dimensions,
as presented in Ref. [24].

Considering the accuracy of the WMAP data which is up to two or three
decimal integers \cite{{26},{27}}, we conclude that the effects of
space-space noncommutative geometry on the spectral index and
its running cannot be currently detected
while the effects of time variability of spatial dimensions
can be detected by the present experimental celectial or ground
instruments. 

The final point that must be emphasized is about decrumpling model
in cosmology. The original motivation of this model presented in
the pioneer paper \cite{1} was based on an {\it ad hoc} assumption
inspired from polymer physics. It is quite possible that this part
of decrumpling model should be revised. However, just how this should
be done is far from obvious. The progress in decrumpling model can only
be made if there is a breakthrough in terms of finding a natural
mechanism for varying the spatial dimension in some alternative
fashion to that which we have considered. 

\section*{Acknowledgments}

It is a pleasure to thank Malcolm MacCallum, Michael Murphy and
Jean-Philippe Uzan for helpful comments.
F.N. thanks Amir and Shahrokh for useful helps.

\end{document}